\documentclass{revtex4}

\usepackage{graphicx}
\newcommand{\sect}[1]{\section{#1}\setcounter{equation}{0}}
\newcommand{\roughly}[1]{\raise.3ex\hbox{$#1$\kern-.75em
\lower1ex\hbox{$\sim$}}}

\def\centeron#1#2{{\setbox0=\hbox{#1}\setbox1=\hbox{#2}\ifdim
\wd1>\wd0\kern.5\wd1\kern-.5\wd0\fi
\copy0\kern-.5\wd0\kern-.5\wd1\copy1\ifdim\wd0>\wd1
\kern.5\wd0\kern-.5\wd1\fi}}
\def\ltap{\;\centeron{\raise.35ex\hbox{$<$}}{\lower.65ex\hbox{$\sim$}}\;}
\def\gtap{\;\centeron{\raise.35ex\hbox{$>$}}{\lower.65ex\hbox{$\sim$}}\;}
\def\gsim{\mathrel{\gtap}}

\def\beq{\begin{equation}}
\def\eq{\end{equation}}
\def\eeq{\end{equation}}
\def\sleptonR{\tilde{\ell}_R}

\def\bino{\widetilde{B}}
\newcommand{\beqa}{\begin{eqnarray}}
\newcommand{\eeqa}{\end{eqnarray}}

\newcommand{\sel}{{\mbox{$\tilde e$}}}
\newcommand{\smu}{{\mbox{$\tilde \mu$}}}

\newcommand{\sell}{{\mbox{$\tilde \ell$}}}

\begin{document}
\bibliographystyle{revtex}


\title{Slepton Flavor Physics at Linear Colliders}



\author{Michael Dine}
\email{dine@scipp.ucsc.edu}
\affiliation{Santa Cruz Institute for Particle Physics,
University of California, Santa Cruz, CA  95064}
\author{Yuval Grossman}
\email{yuvalg@physics.technion.ac.il}
\affiliation{Department of Physics, Technion, Technion City, 32000
Haifa, Israel}
\author{Scott Thomas}
\email{sthomas@stanford.edu}
\affiliation{Department of Physics, Stanford University, Stanford,
CA   94305}


\begin{abstract}
If low energy supersymmetry is realized in nature it is possible that a first
generation linear collider will only have access to some of the
superpartners with electroweak quantum numbers.
Among these, sleptons can provide sensitive probes for lepton
flavor violation through potentially dramatic lepton violating signals.
Theoretical proposals to understand the absence of low energy
quark and lepton flavor
changing neutral currents are surveyed and many are found to predict
observable slepton flavor violating signals at linear colliders.
The observation or absence of such sflavor violation will thus
provide important indirect clues to very high energy physics.
Previous analyses of slepton flavor oscillations are also extended to include
the effects of finite width and mass differences.
\end{abstract}

\maketitle



%
%


\sect{Introduction}

If supersymmetry is discovered at the Tevatron or LHC, much of the
focus of particle physics research will turn to measuring and
understanding the $105$ (or more) soft breaking parameters.  While
this problem may seem, at first, more daunting than understanding
the pattern of quark and lepton masses and mixings, we know, a
priori, that these parameters must show striking regularities.
Otherwise, supersymmetry would lead to large, unobserved flavor
violation among ordinary quarks and leptons.

{}From the perspective of string theory, for example, the prospect
of discovering supersymmetry and measuring the soft breaking
parameters is extremely exciting.   Indeed, in most pictures of
supersymmetry breaking, squark and slepton masses arise from
non-renormalizable operators associated with some new scale of
physics.  This could well be, for example, the scale of string or
M-theory.  Thus, collider measurements of the soft breaking
parameters are potentially indirect probes of extremely high energy
physics, physics we might hope to eventually unravel!

To date, there are only a few theoretical proposals for the
origin of the requisite regularities of the soft parameters
which lead to acceptably small low energy flavor violation among the
quarks and leptons.
Each has important implications for the physics
discoveries of future colliders.  In this note, we will survey
these proposals, and discuss some of their implications for
collider phenomenology.  One
particularly exciting possibility,
is the existence of dramatic flavor
violating phenomena in slepton production.
This can occur even given the strong bounds from
the non-observation to date of flavor violation in low energy
lepton decays.

We also call attention to previous work on slepton flavor violating
collider phenomenology
\cite{Kras,halloscillations} and extend it to include the effects of finite
width and mass differences.
Our motivation is the realization that, in a first phase,
a linear collider is likely to have a center of mass energy of $500$ GeV.
As a result, one can well imagine that supersymmetry will have been
discovered at the LHC, but that we will have only limited information
about the spectrum, and a first generation linear collider will
have access to only a few of the lightest superpartner states.
In many theories these include sleptons.  Even with this limited set of
states, the observation -- or not -- of slepton flavor violation will
provide important clues to the underlying supersymmetry breaking
structure.

The reason that observable sflavor violation at a linear collider
is possible is easy to
understand.  In many proposals to solve the supersymmetric flavor problem,
a high degree of degeneracy among sleptons (and squarks) is predicted.
As a result,
there is the potential for substantial mixing of flavor
eigenstates.
This can lead to substantial and observable
sflavor violation.
To be readily observable, it is necessary that mass splittings
between the states not be too much smaller than the decay widths, and
that the mixing angles not be terribly small. In some of these
proposals, and in particular gauge mediation, the predicted degree
of degeneracy is extremely high, and the mixing probably
unobservable. However, in most other proposals, the splittings can
be comparable or even larger than the widths, and the mixing
angles may be of order the Cabbibo angle or larger. In this case,
dramatic collider signatures are possible.


\section{Mediation Mechanisms}

It is instructive to review the various proposed mechanisms
for understanding the suppression of flavor changing processes in
supersymmetric theories.  We should start by noting that most
analyses of collider experiments work in the framework of
what has become to known as Gravity
Mediation.  In such models, one simply assumes exact
degeneracy of squarks and sleptons at some very high energy scale.
This assumption is not natural since it is violated by quantum
corrections (and is in fact scale dependent),
and so surely breaks down at some level. Without
some theory, or more detailed assumptions, one cannot assess
the degree to which this assumption is viable.  In other words,
degeneracy in this case
is a puzzle to be explained rather than a mechanism in itself.

There are a number of more serious proposals of which
we are aware for understanding the suppression
of supersymmetric contributions to quark and lepton flavor
violating processes.
These fall into three broad classes.
The first are mechanisms which seek to ensure
a high degree of degeneracy as the result of dynamics
without specific assumptions about flavor.
The second are flavor symmetries which enforce either
a high degree of degeneracy or alignment of the squark and
slepton mass matrices with those of the quarks and leptons.
The third invokes heavy superpartner masses to kinematically
suppress low energy processes.
\begin{itemize}
\item  Gauge Mediation:  In its simplest form, any flavor
violation in gauge mediation occurs at a high order in the loop
expansion,
or due to non-renormalizable operators.
In the former case, the
suppressions are typically by powers of small Yukawa couplings and
extra loop factors.
These effects are automatically aligned with flavor violation
in the Yukawa sector and are therefore not dangerous and do
not give rise to interesting processes.
In the latter case, the suppression is by powers
of $\Lambda^2/M^2$, where $\Lambda$ represents some typical scale
associated with the supersymmetry-breaking interactions, and $M$
some large scale associated with new physics such as
a flavor scale at which flavor is spontaneously broken or the Planck or
string scale.
For $\Lambda \ll M$ the distinctive feature is the absence of
sflavor violating processes.
\item  Dilaton Domination:  At weak coupling in the heterotic
string
there is a regime in which
a gravity-mediated spectrum with a high degree of degeneracy
is obtained.
If the weak coupling
picture is valid, one might hope that generic flavor violating corrections
to squark and slepton mass matrices are of order
${\alpha_{\rm GUT} / \pi}$.
Numerically this is
just enough to understand the suppression of
flavor changing neutral currents  \cite{louisnir}.
It suggests that mass splittings among the lightest sleptons will be at
least of order a few parts in $10^{-3}$.
We will see that this is within the range
(albeit at the low end) of what
one might hope to measure directly at a linear collider.
Without further
assumptions about flavor, one expects mixings of squarks and
sleptons to be of order one.
Such a mass splitting is
comparable to the expected decay widths of these states, so this
sflavor violating
mixing might well be observable.
In the
strongly coupled limit, it seems likely that the violations of
degeneracy are larger \cite{adgt}.
\item  Anomaly Mediation:  The anomaly-mediated hypothesis
superficially has
some features in common with gauge mediation.
However, the mediation scale
is now comparable to the Planck scale, and one has to ask about
the magnitude of flavor violating corrections to squark and
slepton mass matrices.
It has been argued that brane world realizations of supersymmetry
breaking might naturally provide a context for anomaly
mediation with small corrections to degeneracy \cite{rs}.
However, anomaly mediation turns out not a robust feature of
brane world supersymmetry breaking, and violations of degeneracy
are generally large \cite{adgt}.  Still, the
degeneracy might be small enough to suppress low energy flavor
violating processes in very special
circumstances.
\item  Gaugino Domination (and the closely
related idea of gaugino mediation):
With no-scale boundary conditions one
assumes that scalar masses vanish or are very small compared to gaugino
masses at the high messenger scale.
Without a detailed underlying model, it is hard to
know how large the violations of degeneracy might be at this
scale, but it seems
reasonable to suppose that at the high scale, the
magnitude of the squark and slepton
mass matrices are suppressed relative to gaugino masses by an amount of
order $\alpha_{\rm GUT} / \pi$.
At lower scales, renormalization
group evolution gives a flavor independent
gaugino mass contribution to slepton masses which leads
to a high degree of degeneracy.
As in the case of dilaton
domination, without further assumptions about flavor, the violations
of degeneracy are expected to be maximally
flavor violating, and to lead
to large mixings in the high scale squark and slepton
masses which in turn leads to large mixings at the electroweak
scale.
\item Conformal Sequestering:
Another possibility for obtaining a high degree of
degeneracy is to postulate that the first two
generation squarks and sleptons are coupled to an approximately
conformal sector over a few orders of magnitude in renormalization
group evolution \cite{nelsonstrassler}.
This has the effect of exponentially suppressing
squark and slepton masses as well as fermion Yukawa couplings.
Below this approximately conformal range of scales further renormalization
group evolution gives a flavor independent gaugino
mass contribution to slepton masses which leads to a high
degree of degeneracy much as with no-scale boundary conditions.
Violations of conformal invariance by standard model gauge interactions
limit the degree of slepton degeneracy to roughly
$\alpha_{\rm W} /  \pi \sim 10^{-2}$.
Without further assumptions about flavor in the high scale
theory above the approximately conformal scale, slepton
mixings are related to ratios of lepton masses by roughly
$\sin \phi_{ij} \gsim
\sqrt{m_{\ell_i} / m_{\ell_j}}$.
\item  Non-Abelian Flavor Symmetries:  If the
explanation of squark and slepton degeneracy lies in non-abelian
flavor symmetries, it is reasonable to expect that violations of
degeneracy are correlated with the values of quark and lepton
masses and the KM angles.  In this case, one might hope to obtain
tighter predictions in a given model
for the pattern of flavor violation in the slepton sector.
The level of degeneracy is
model dependent, but again a few parts times $10^{-3}$ is a reasonable
expectation for the violations of degeneracy, with values for the
mixings of order Cabbibo angles $\sin \phi_{ij} \sim
\sqrt{m_{\ell_i} / m_{\ell_j}}$ \cite{nonabelianflavor}.
\item  Abelian Flavor Symmetries:  As an alternative to degeneracy
among squarks and sleptons, it has been suggested that the squark
and slepton mass matrices might be approximately aligned with the
quark and lepton matrices \cite{nirseiberg}. This can come about in
theories with Abelian (discrete) flavor symmetries.  In this case,
one does not expect any approximate degeneracy among sleptons.  
In order to suppress flavor changing lepton decays
the mixings need to be somewhat small. 
Still, the mixing can be large enough, in particular for mixing involving taus, 
such that sflavor violation can be observable at colliders.
\item First Two Generations Heavy:
Another possibility to suppress dangerous levels of
supersymmetric contributions to low energy
quark and lepton flavor violation is to postulate that
the superpartners are very heavy.
The first two generation squarks and sleptons can have
masses up to of order 20 TeV without introducing
significant tuning of electroweak symmetry breaking \cite{ckn}.
In this case only the (mostly) stau slepton(s) would be kinematically
accessible at future colliders.
Without additional assumptions about flavor, naturalness
of the full slepton mass matrix implies that
mixing of this light state(s) among the flavor eigenstates
would be of order $m/M$ where $m$ and $M$ are the light
and heavy slepton masses.
\end{itemize}

In sum, of the various proposals to understand the absence of
flavor violation at low energies, several predict dramatic violations of flavor
at colliders.  Among those which don't, there tend to be
distinctive predictions for the spectrum.  Models of gauge
mediation, for example, tend to be highly predictive.
While there is no one compelling
model of this type, many models exist, and one can imagine
detailed measurements distinguishing between them.
In the case of alignment mechanisms, while there should flavor
mixing it will not be so dramatic, one should observe correlations
between the squark and slepton and the ordinary quark and lepton
masses.

In the case of
non-Abelian flavor symmetries, not only does one
expect significant mixing, but one can hope to obtain, given some
assumptions about the form of flavor symmetry breaking, precise
predictions for some of the violations of flavor symmetry.
Moreover, these are likely to be correlated with quark, lepton and
neutrino mass matrices. In such a case, precision measurements
might ultimately permit distinguishing between different models.
Clearly, all of these are directions for further theoretical work,
but the discovery of supersymmetry and unraveling the pattern of
symmetry breaking would provide important insights into the nature
of physics at very high energy scales.
The observation of flavor violation -- or its absence -- at linear
colliders will provide important clues to the nature of the
underlying mechanisms of supersymmetry breaking and mediation.

\sect{Sflavor Oscillations of Unstable Sleptons}
Slepton flavor oscillations arise if the sleptons mass eigenstates are
not flavor eigenstates.  Consider for simplicity the case in which
leptons are produced initially in flavor eigenstates.  If the mass
eigenstates have distinct mass, then the flavor eigenstate oscillates
in time and space with a frequency given by the mass splittings.  If
the slepton decay rate is much smaller than the mass splitting,
oscillations average out and the probability of decay from a given
flavor eigenstate is given simply by mixing angles, as implicitly
assumed previously \cite{Kras}.  If the slepton decay is rapid
compared with the oscillation frequency then the probability of decay
to another flavor eigenstate is suppressed \cite{halloscillations}.
There are additional effects which can affect the flavor violating
decay probability.  First, different flavor eigenstates need not have
the same decay width.  This is particularly true for the
$\tilde{\tau}$ slepton which can have a non-trivial decay amplitude to
the Higgsino component of a neutralino at moderate to large $\tan
\beta$.  In addition, the probability amplitude or cross section for
production of different mass eigenstates need not be equal.  This is
potentially important in $S$-wave processes such as $e^- e^-
\rightarrow \tilde{\ell}^- \tilde{\ell}^-$ near threshold due to
finite mass differences.  Both of these effects might in principle
enhance flavor violating effects in decays, and are considered below.

For simplicity throughout we consider the two flavor CP conserving case
with flavor and mass eigenstates related by
\beq |\sel\rangle = c |1\rangle +s|2\rangle, \qquad
|\smu\rangle = -s |1\rangle+c|2\rangle
\eeq
where $c$ and $s$ are complex numbers that satisfy
$|c|^2+|s|^2=1$. (Only in the limit where the width difference is
neglected $s$ and $c$ can be chosen to be real, thus getting they regular
interpretations as a sine and a cosine of an angle.)
The other relevant physical dimensionless parameters
characterize the ratio of mass splitting or oscillation
frequency to decay width,
the relative width difference of the mass eigenstates,
and the relative rate or cross section for production of the
mass eigenstates
\beq x \equiv {\Delta
m \over \Gamma}, \qquad y \equiv {\Delta \Gamma \over 2\Gamma},
\qquad z \equiv {\Delta \sigma \over 2\sigma}
\eeq
where
$\Delta m$ and $\Delta \Gamma$
are the mass and width differences of the
two mass eigenstates, and $\Gamma$ is their average width.
$\sigma_i$ is the cross section to produce the $i$th mass
eigenstate, and
\beq \sigma={\sigma_1+\sigma_2\over
2}, \qquad \Delta \sigma=\sigma_2-\sigma_1.
\eeq
The case $x=y=z=0$ was considered in \cite{Kras} while
$y=z=0$ with $x$ arbitrary was considered in
\cite{halloscillations}.
In the first subsection below, the flavor decay probability is
derived for small but non-vanishing
$x$, $y$, and $z$.
In the next subsection the decay probability for
small $x$ and $y$ is shown to depend directly on
the off-diagonal slepton mass squared mixing, as required
in this limit. In the final subsection,
estimates of the widths are given.

\subsection{Decay Probability}

The probability for an initial selectron state
to decay as a smuon state, $P(\sel \to \smu)$,
may be calculated from the time evolution of the state.
A selectron flavor state produced at time $t=0$
is a linear combination of mass eigenstates
\beq
\psi(t=0) =N\left[\sigma_1 c |1\rangle +
\sigma_2 s|2\rangle\right] = N\left[\sigma|\sel\rangle
-\Delta\sigma|\smu\rangle \right]\,,
\eeq
where the normalization factor is
\beq N=\left(c^2\sigma_1^2 + s^2
\sigma_2^2\right)^{-1/2}
\eeq
and $\sigma_i \equiv \sigma_i/\sigma$ are the dimensionless
cross sections.
The factors of $\sigma_i$ in the relative amplitudes
account for the possibility that the cross section for
each mass eigenstate is distinct; for example, from mass
difference effects near threshold.
Using the standard oscillation
formalism the time evolution formula for a
selectron initial state is
\beq
\psi(t) =N\left[ c \sigma_1
e^{-i\mu_1 t}|1\rangle + s \sigma_2  e^{-i\mu_2
t}|2\rangle\right]\,,
\eeq
where
\beq
\mu_i = m_i -i\Gamma_i/2
\eeq
and $m_i$ ($\Gamma_i$) is the mass (width) of the $i$th
mass eigenstate.
The time dependent oscillation probability is given by
\beq
P(\sel \to \smu)[t] = {\left|\langle
\smu|\psi(t)\rangle\right|^2 \over \left|\langle
\psi(t)|\psi(t)\rangle\right|^2}
\eeq
so that the projection onto the smuon flavor eigenstate is
\beq
\left|\langle \smu|\psi(t)\rangle\right|^2 = N^2
c^2 s^2 \left[\sigma_2^2 e^{-\Gamma_2 t} + \sigma_1^2 e^{-\Gamma_1
t} - 2  \sigma_1 \sigma_2 e^{-\Gamma t} \cos (\Delta m t)
\right]\,.
\eeq
After integration over time from zero to infinity
\beq
\int_0^{\infty}dt
\left|\langle \smu|\psi(t)\rangle\right|^2 = N^2 c^2
s^2\left[ {\sigma_2^2 \over \Gamma_2} +{\sigma_1^2 \over \Gamma_1}
- {2\sigma_1 \sigma_2 \Gamma\over \Delta M^2 + \Gamma^2}\right]\,,
\eeq
and
\beq
\int_0^{\infty} dt
\left|\langle \psi(t)|\psi(t)\rangle\right|^2=
N^4\left[ {s^4\sigma_2^4 \over \Gamma_2} +{c^4\sigma_1^4 \over
\Gamma_1} + {2c^2 s^2\sigma_1^2 \sigma_2^2 \over  \Gamma}\right]\,.
\eeq
Then the time integrated oscillation probability is
\beq
P(\sel \to \smu) = {|cs|^2 \over N^2}\left(
{\sigma_2^2 \over \Gamma_2} +{\sigma_1^2 \over \Gamma_1} -
{2\sigma_1 \sigma_2 \Gamma\over \Delta M^2 + \Gamma^2}\right)
\times
\left(
{|s|^4\sigma_2^4 \over \Gamma_2} +{|c|^4\sigma_1^4 \over \Gamma_1} +
{2|cs|^2\sigma_1^2 \sigma_2^2 \over  \Gamma}\right)^{-1}.
\eeq
%
It is useful to expand
$P$ in the (presumably) small parameters, $x$, $y$ and $z$. To
lowest order
\beq
P(\sel \to \smu) =2 c^2 s^2 \left[x^2
+y^2 + 2z^2 - 2 z y\right]
\label{probeq}
\eeq
The oscillation probability in this limit is
quadratic in all three small parameters.
The $x^2$ term describing the effect of finite mass
difference is the lowest order
form of the effect discussed in \cite{halloscillations}.
The $y^2$ term describing the effect of a width difference between
the two mass eigenstates parametrically plays a similar
role as $x^2$, as is well known from the $D$-meson
system.
The $z$ parameter describes the
effect of the difference of production probability for the two
mass eigenstates. This effect has
never been mentioned in the context of meson
mixing, since there it is always completely negligible. As
discussed below, also in the present case of slepton production,
its effect is somewhat smaller than the
effects of $x$ and $y$.


We first consider the possible effect of the $z$ parameter.
Ignoring the decay width, the cross
section for $e^- e^- \rightarrow \sell^- \sell^-$ is $S$-wave
and proportional to the final state velocity $\beta$ near
threshold.
Far above threshold, where the cross
section is large, $\beta \to 1$ and $z$ vanishes.
Only near threshold, where the cross section is small, $z$ can be sizeable.
The Monte Carlo program Pandora may be used to estimate $z$.
As an example, we considered small mixing angles and took
$m_{\sell}=150$ GeV, $\Delta m=1$ GeV, $m_{\chi^0}=100$
and a center of mass energy of 305 GeV, and found $z=0.13$. Note that in
this case $\Gamma=240$ MeV, so $x \approx 4$, and it is the
dominant effect.
Also at smaller $x$ the inequality $z < x$ still holds.
We conclude that the effect of $z$ is smaller than that of $x$ and $y$.
This is because $z$ is generated only for finite $x$ or $y$,
and in such a way that it is smaller then $x$ and $y$.
Even so, it does increase the flavor violating decay
probability somewhat.


\subsection{Calculating $\lowercase{x}$ and $\lowercase{y}$}

It is instructive to derive the $x$ and $y$ parameters
from the general form of the slepton mass matrix
including the effects of flavor mixing and decay
\beq
M=\pmatrix{m_{11}-i\Gamma_{11}/2 & m_{12}-i\Gamma_{12}/2
\cr m_{21}-i\Gamma_{21}/2 & m_{22}-i\Gamma_{22}/2}
\eeq
where $m_{21}=m_{12}^*$, $\Gamma_{21}=\Gamma_{12}^*$ and $m_{ii}$ and
$\Gamma_{ii}$ are real.
Solving the eigenvalue equation
$\det(M-\mu_i)=0$ gives the mass eigenstates, the mass and
width differences, and the mixing angles.
The general calculation
is not very illuminating, so several simplifying assumptions
are employed.

First, we note that in the MSSM there are no flavor violating
decays. Thus, we set $\Gamma_{21}=0$. (This may not be the case in
models beyond the MSSM, for example, in supersymmetric models with
4 Higgs doublets, where the Higgs bosons can mediate flavor
changing decays.) Second we assume CP conservation
and then  $m_{12}$ may be taken real.
Next, for simplicity, we assume that
$m_{11}-m_{22}\ll m_{12}$ and may be
approximated by zero.
This is because in the opposite limit $m_{11}-m_{22}\gg
m_{12}$ the mixing angle is small and we are not interested in
that case.

With these assumptions we solve the eigenvalue equation.
We define the real parameters
$a \equiv m_{11}=m_{22}$, $b=\Gamma_{11}/2$, $d=\Gamma_{22}/2$ and
$f=m_{12}$ and find that
for  $(b-d)>2f$
\beq
x=0,\qquad y={\sqrt{(b-d)^2-4 f^2}\over b+d},
\eeq
while for $(b-d)<2f$
\beq
x={2\sqrt{4 f^2-(b-d)^2} \over b+d},\qquad  y=0.
\eeq
For simplicity for the mixing angles we present results only
in some limits.
For $(b-d)\gg2f$ we find
\beq
c\approx {1\over \sqrt{2}}(1+i), \qquad
s\approx {f \over \sqrt{2} (b-d)} (1-i),
\eeq
and for $(b-d)\ll 2f$ we find
\beq
c \approx s \approx {1\over \sqrt{2}}  \;.
\eeq
We see that in the $(b-d)\gg2f$ limit, where the width
difference is large, the mixing angles are suppressed.
Actually, the oscillation probability is the same in both limits
considered above
\beq
P\approx 2|cs|^2 (x^2+y^2) \approx {2 m_{12}^2 \over \Gamma^2}\;.
\eeq
In particular, it does not depend on $\Gamma_{11}-\Gamma_{22}$.

We conclude
that no matter if we have large width difference or not, the
oscillation probability is determined by the ratio between the
off-diagonal mass term and the average width. While we made several
assumptions in order to derive the above results, the conclusion is general.
Off-diagonal mixing terms in the mass squared matrix must
not be too much smaller than the decay widths
in order to obtain large violations of flavor in slepton decays.

\subsection{Decay Width}

As we saw the relevant parameters that determined the flavor violation
oscillation probability are $M_{12}/\Gamma$ and to some extent also
$(\Gamma_{11}-\Gamma_{22})/\Gamma$.
While $M_{12}$, the off diagonal mass difference,
is a parameter of the model, the widths
are derived from the model parameters.
Therefore, below we estimates their sizes.

We start with estimating $\Gamma$.
For example we consider decays of right handed sleptons.
The decay width for a right handed slepton to decay to a Bino
through the emission of a lepton is
\beq
\Gamma( \sleptonR \rightarrow \ell_R \bino) =
{ \alpha ~m_{\sleptonR} \over 2~\cos^2 \theta_W}
\left( 1 - { m_{\bino}^2 \over m_{\sleptonR}^2 } \right)^2
\eeq
where $\alpha$ is the fine structure constant evaluated
at the slepton mass scale, $\theta_W$ is the weak mixing angle,
and the lepton mass has been ignored.
Numerically, the dimensionless width is
\beq
{1 \over m_{\sleptonR}} \Gamma( \sleptonR \rightarrow \ell_R \bino)
\simeq
5 \times 10^{-3} \; \left(1 -  m_{\bino}^2 \over m_{\sleptonR}^2\right)^2
\eeq
For a typical value of  $m_{\bino} / m_{\sleptonR}= 0.75$
the dimensionless width is
$\Gamma( \sleptonR \rightarrow \ell_R \bino)/m_{\sleptonR} \simeq
10^{-3}$.

In light of our remarks about the expected degrees of degeneracy
in various proposals for supersymmetry breaking, this is a
striking number.  In some sense, within the range of ideas which
have been proposed, a large fraction predict that the splittings
should  be as large or larger than the width, and thus observable.

Next we consider
the possible magnitude of $\Gamma_{11}-\Gamma_{22}$ for slepton
production and decay by $\sell \rightarrow \chi \ell$.
If $m_{\sell}- m_\chi^0 \gg m_\tau$, the final
state neutralino is pure
gaugino, and the sleptons are pure left- or right-handed eigenstates, then
the decay is universal and  $\Gamma_{11}=\Gamma_{22}$.
Universality violation due
to phase space can be significant only when the mass splitting
between the sleptons and the neutralino is at least comparable to or
smaller than the lepton mass.
Another
effect can be dues to significant Higgsino component in $\chi$.
The full calculation for $\sell \rightarrow \chi
\ell$ for general gaugino--Higgsino mixing can be found in
\cite{nijiri}.  A very crude estimate
in the the $\tilde{\tau}-\tilde{\mu}$ case with large $\tan \beta$ is
\beq
{\Gamma_{11}-\Gamma_{22}\over \Gamma}
 \sim {Y_\tau^2 \over Y_\tau^2 + 2 g_1^2 Z_{1B}^2 / Z_{1h}^2 }
\eeq
where $Z_{1h}$ and $Z_{1B}$ are the Higgsino and Bino amplitudes
of $\chi$, $Y_\tau$ is the $\tau$ Yukawa coupling, and $g_1$ is the
hypercharge coupling.  For large $\tan\beta$ and significant Higgsino
component in the neutralino $\Gamma_{11}-\Gamma_{22}$ can be
large.

\sect{Conclusions}

If superpartners are discovered, there is good reason to
expect that
sleptons will have a high degree of degeneracy which
enhances sensitivity to sflavor violating effects.
It is quite possible
then that significant sflavor violation could be observed
at a linear collider.
The observation -- or non-observation -- of such
sflavor breaking would provide important indirect clues about
flavor and supersymmetry physics at potentially
extremely high energy scales.

\subsection*{Acknowledgements}
Y.G. Thanks the SLAC theory group for their warm hospitality
while most  of this work were done. The work of M.D. was supported
in part by a grant from the US Department of Energy. The work of
Y.G. was supported in part by the Israel Science Foundation under
Grant No.~237/01-1, and in part by the United States--Israel
Binational Science Foundation through Grant No. 2000133. The work
of S.T. was supported in part by the US National Science
Foundation under grant PHY98-70115, the Alfred P. Sloan
Foundation, and Stanford University through the Fredrick E. Terman
Fellowship.


\end{document}